# Electric-field-controlled altermagnetic spin-momentum locking and ultrahigh tunneling magnetoresistance in bilayer $Cr_2SeO$

Yunfei Gao[1], Aolin Li[1,*], Zesen Fu[1], Bei Zhang[1], Haiming Duan[1], Fangping Ouyang[1,2,3,4,†]

[1] School of Physical Science and Technology, Xinjiang Key Laboratory of Solid-State Physics and Devices, Xinjiang University, Urumqi 830046, People's Republic of China

[2] School of Materials and Chemical Engineering, Xinjiang Polytechnic Institute, Hotan Xinjiang, 848000, P.R. China

[3] School of Physics, Institute of Quantum Physics, Hunan Key Laboratory for Super-Microstructure and Ultrafast Process, and Hunan Key Laboratory of Nanophononics and Devices, Central South University, Changsha 410083, People's Republic of China

[4] State Key Laboratory of Powder Metallurgy, and Powder Metallurgy Research Institute, Central South University, Changsha 410083, People's Republic of China

[*]Contact author: liaolin628@xju.edu.cn

[†]Contact author: ouyangfp06@tsinghua.org.cn

Altermagnets offer a promising platform for spintronics by combining zero net magnetization with nonrelativistic spin splitting, yet efficient electrical manipulation of their spin-dependent transport remains challenging. Here, using first-principles calculations, we show that an out-of-plane electric field induces spin-momentum locking in bilayer $Cr_2SeO$, gives rise to a magneto-optical Kerr effect, and drives a semiconductor-to-metal transition at sufficiently high field magnitudes. Reversing the field direction switches the spin-momentum-locking pattern, which reverses the sign of the Kerr rotation angle as well as that of the longitudinal and transverse spin conductivities. Leveraging this electric-field-controlled spin-momentum locking, we propose a dual-gate altermagnetic tunnel junction in which high- and low-resistance states are realized without reversing the Néel vector. The resistance switching originates from the matching and mismatching of spin-polarized conducting channels in momentum space, yielding a tunneling magnetoresistance ratio on the order of $10^8$. Our findings provide a theoretical basis for realizing electric-field-controlled spin-momentum locking in bilayer $Cr_2SeO$ and developing all-electrically controlled altermagnetic spintronic devices.

## I. INTRODUCTION

Altermagnets, an emerging class of magnetic materials, have attracted extensive interest in both fundamental research and device applications [1-4]. They combine the zero net magnetization of conventional antiferromagnets with the spin-split bands of ferromagnets. The splitting is nonrelativistic and momentum dependent, alternating in sign across the Brillouin zone, which gives rise to a characteristic spin-momentum locking. A growing number of altermagnets have been identified, including bulk

MnTe and CrSb as well as two-dimensional (2D) $Fe_2Se_2O$, $Ca(CoN)_2$, $V_2SSeO$, and $Cr_2SeO$ [5-12]. The distinctive electronic and symmetry properties make altermagnets promising for spintronic applications [7,8,13-17]. For example, altermagnets can exhibit a finite magneto-optical Kerr effect (MOKE) when permitted by their magnetic symmetry [18-20]. They can also exhibit pronounced anisotropy in spin-dependent transport, with the resulting spin current depending on the direction of the applied electric field relative to the crystal axes [1,21].

Magnetic tunnel junctions (MTJs) play a pivotal role in spintronic applications [22-32]. A conventional MTJ typically consists of a ferromagnet/tunneling barrier/ferromagnet sandwich structure. By applying external magnetic fields or current-induced spin torques, such as spin-transfer torque (STT) and spin-orbit torque (SOT), the magnetization directions of the two ferromagnetic electrodes can be switched between a parallel configuration (PC) and an antiparallel configuration (APC), thereby modulating the MTJ between low- and high-resistance states [33]. However, conventional MTJs are susceptible to stray fields from the ferromagnetic layers as well as external magnetic perturbations, which limits their stability and high-density integration at the nanoscale.

Altermagnets provide an ideal solution to this issue. They intrinsically possess the robustness against stray fields and external magnetic perturbations characteristic of antiferromagnets, while their spin-momentum locking allows for PC and APC states to achieve distinct resistance switching. Consequently, altermagnetic tunnel junctions (AMTJs) have recently been proposed as a promising route for advanced spintronic devices [13,34-38]. Most reported AMTJs rely on Néel-vector reversal to switch between high- and low-resistance states. In practical implementations, such reversal is commonly achieved through current-driven manipulation [39,40]. An alternative strategy is to electrically reconfigure the spin-polarized electronic states while keeping the magnetic order unchanged. Recent studies have predicted electric-field-controlled spin-momentum locking in bilayer $Cr_2SO$ and $Cr_2SeO$ [41,42], suggesting that the spin-polarized conducting channels of these materials can be manipulated directly by an out-of-plane electric field. However, the magneto-optical and spin-transport functionalities enabled by such electrically switchable spin textures remain largely unexplored.

In this work, we investigate the magneto-optical and transport consequences of electric-field-controlled spin-momentum locking in bilayer $Cr_2SeO$. Density functional theory (DFT) calculations show that an out-of-plane electric field breaks the $PT$ symmetry of the bilayer, thereby lifting the spin degeneracy, inducing spin-momentum locking, and permitting finite magneto-optical and spin-dependent transport responses. Reversing the field switches the spin-momentum-locking pattern and reverses the signs of the Kerr rotation and the strongly direction-dependent longitudinal and transverse spin conductivities. We further exploit this electrical switchability to design a dual-gate bilayer $Cr_2SeO$ AMTJ without reversing the Néel vector in either electrode. By independently tuning the gate voltages applied to the two electrodes, the spin-polarized conducting channels can be either matched or mismatched in momentum space. Quantum transport calculations using the DFT-based nonequilibrium Green's function formalism (DFT-NEGF) reveal that the proposed AMTJ exhibits an ultrahigh tunneling magnetoresistance (TMR) ratio on the order of $10^8$.

## II. CALCULATION METHOD

The structural optimization and electronic structure calculations were carried out using the Vienna ab initio simulation package (VASP) based on DFT [43], where the exchange-correlation interaction was treated with the generalized gradient approximation (GGA) of the Perdew-Burke-Ernzerhof (PBE) form. The projector augmented wave (PAW) pseudopotentials were employed and the energy cutoff was set to 500 eV [44,45]. A Γ-centered 17 × 17 × 1 k-point mesh was used for Brillouin zone integration. The force criterion for structural optimization and the energy criterion were set to be 0.01 eV/Å and $10^{-6}$ eV, respectively. To properly deal with the strong correlation effect in Cr atoms, the DFT + U method was adopted with an effective Coulomb parameter U = 3.55 eV for 3d orbital [4]. A vacuum layer thicker than 15 Å was applied to avoid interactions between periodic images and the van der Waals interactions were considered using the DFT-D3 method of Grimme [46]. The phonon dispersions were calculated using the finite-displacement method as implemented in VASP, with a 3 × 3 × 1 supercell. The optical and spin-resolved charge conductivities were calculated using Wannier-interpolated tight-binding Hamiltonians constructed with the Wannier90 package [47]. The optical conductivity was obtained using the Kubo-Greenwood formula with spin-orbit coupling (SOC) on a dense 300 × 300 × 1 k-point mesh [48]. The spin-resolved charge conductivity was evaluated within Boltzmann transport theory under the constant relaxation-time approximation [49]. The temperature and relaxation time were set to 300 K and 10 fs, respectively, and a 300 × 300 × 1 k-point mesh was employed.

The electronic transport properties of the AMTJ were calculated using Nanodcal package based on the DFT-NEGF [50], where the exchange-correlation interaction was described by the GGA of the PBE form. The double-zeta polarized (DZP) basis was employed with the energy cutoff of 100 Hartree, and the energy convergence criterion was set to $10^{-5}$ eV for the transport calculations. The k-point mesh of the electrodes was set to 1 × 17 × 100. The spin-dependent transmission $T_\sigma$ as a function of energy $E$ and bias $V$ can be described by the following equation:

$$T_\sigma(E,V) = Tr[\Gamma_L(E,V)G^R(E,V)\Gamma_R(E,V)G^A(E,V)], \quad (1)$$

where $\sigma$ represents spin direction (↑, ↓), $\Gamma_{L/R}$ is the linewidth function, which describes the coupling between the left or right electrodes and the central region. $G^R$ and $G^A$ are the retarded and advanced Green's functions, respectively. $I_\sigma$ is defined by the following Landauer-Büttiker formula:

$$I_\sigma(V) = \frac{e}{h}\int_{\mu_L}^{\mu_R} T_\sigma(E,V)[f(E-\mu_L) - f(E-\mu_R)]dE, \quad (2)$$

where $h$ is Plank's constant, $e$ is the electron charge, $\mu_{L/R} = E_F \pm \frac{eV}{2}$ are the chemical potentials with bias V, and $f(E-\mu_{L/R})$ are the Fermi-Dirac distribution functions of electrons in the two electrodes under non-equilibrium conditions.

## III. RESULTS AND DISCUSSIONS

### A. Electric-field control of spin-momentum locking

Monolayer $Cr_2SeO$ is a square-lattice 2D altermagnet, as illustrated in Fig. 1(a). Structural relaxation yields an in-plane lattice constant of 3.72 Å, in good agreement with the previously reported values [4,12]. The corresponding first Brillouin zone and high-symmetry points are also shown in Fig. 1(a). A Néel temperature of approximately 280 K has been reported for monolayer $Cr_2SeO$, which exhibits easy-plane magnetic anisotropy [12]. We construct bilayer $Cr_2SeO$ by van der Waals stacking, with its lowest-energy magnetic configuration shown in Fig. 1(b). By including SOC, bilayer $Cr_2SeO$ exhibits an in-plane anisotropy, with a magnetocrystalline anisotropy energy (MAE) of 0.52 meV per unit cell. A schematic illustration of the MAE is shown in Fig. 1(c). We then compare the electronic structures of monolayer and bilayer $Cr_2SeO$. As shown in Fig. 1(d), monolayer $Cr_2SeO$ is a semiconductor with a band gap of 0.59 eV and exhibits nonrelativistic momentum-dependent spin splitting. Distinct spin-polarized valleys emerge at the high-symmetry points X and Y. The valence-band maximum and conduction-band minimum exhibit opposite spin polarizations. In contrast, the $PT$ symmetry of bilayer $Cr_2SeO$ enforces spin degeneracy throughout the Brillouin zone. The bilayer $Cr_2SeO$ therefore exhibits spin-degenerate bands while retaining its semiconducting character, with a band gap of 0.59 eV, as shown in Fig. 1(e). Finally, the phonon spectrum in Fig. 1(f) contains no imaginary frequencies, confirming the dynamical stability of bilayer $Cr_2SeO$.

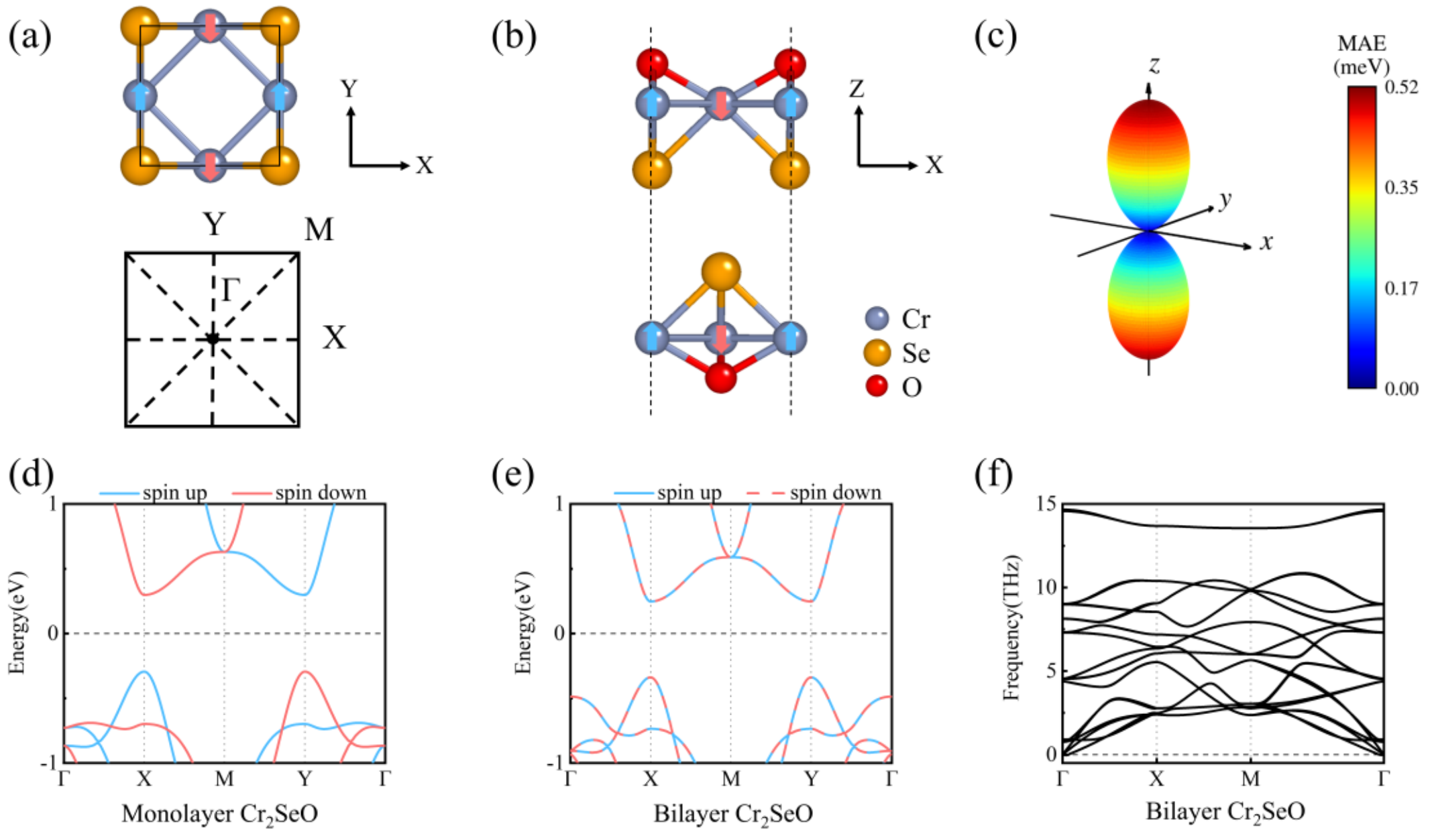

FIG. 1. (a) Top view of monolayer $Cr_2SeO$ and the corresponding first Brillouin zone, with the high-symmetry points indicated. (b) Side view of bilayer $Cr_2SeO$. The red and blue arrows denote opposite orientations of the magnetic moments on the Cr atoms. (c) Schematic three-dimensional MAE including SOC. Spin-resolved band structures of (d) monolayer and (e) bilayer $Cr_2SeO$. The Fermi energy is set to zero. (f) Phonon dispersion of bilayer $Cr_2SeO$.

To modulate the band structure of bilayer $Cr_2SeO$, we apply an out-of-plane electric field. Applying an out-of-plane electric field generates a potential difference between the upper and lower layers, thereby breaking the $PT$ symmetry. This lifts the spin degeneracy and inducing significant spin splitting at the high-symmetry points X and Y, accompanied by a reduction in the band gap. When a positive electric field is applied, the valleys at the high-symmetry point X exhibit spin-up polarization, while those at the high-symmetry point Y show spin-down polarization. Conversely, when a negative electric field is applied, the spin polarizations at X and Y valleys reverse. At critical field strengths of ±0.85 V/Å, the bands at the high-symmetry points X and Y cross the Fermi level, driving an electric-field-induced semiconductor-to-metal transition, as shown in Figs. 2(a) and 2(b). The evolution of the band gaps at the high-symmetry points X and Y as a function of the out-of-plane electric field is presented in Fig. 2(c). These results demonstrate that the spin-momentum locking in bilayer $Cr_2SeO$ can be reversibly controlled by an external out-of-plane electric field.

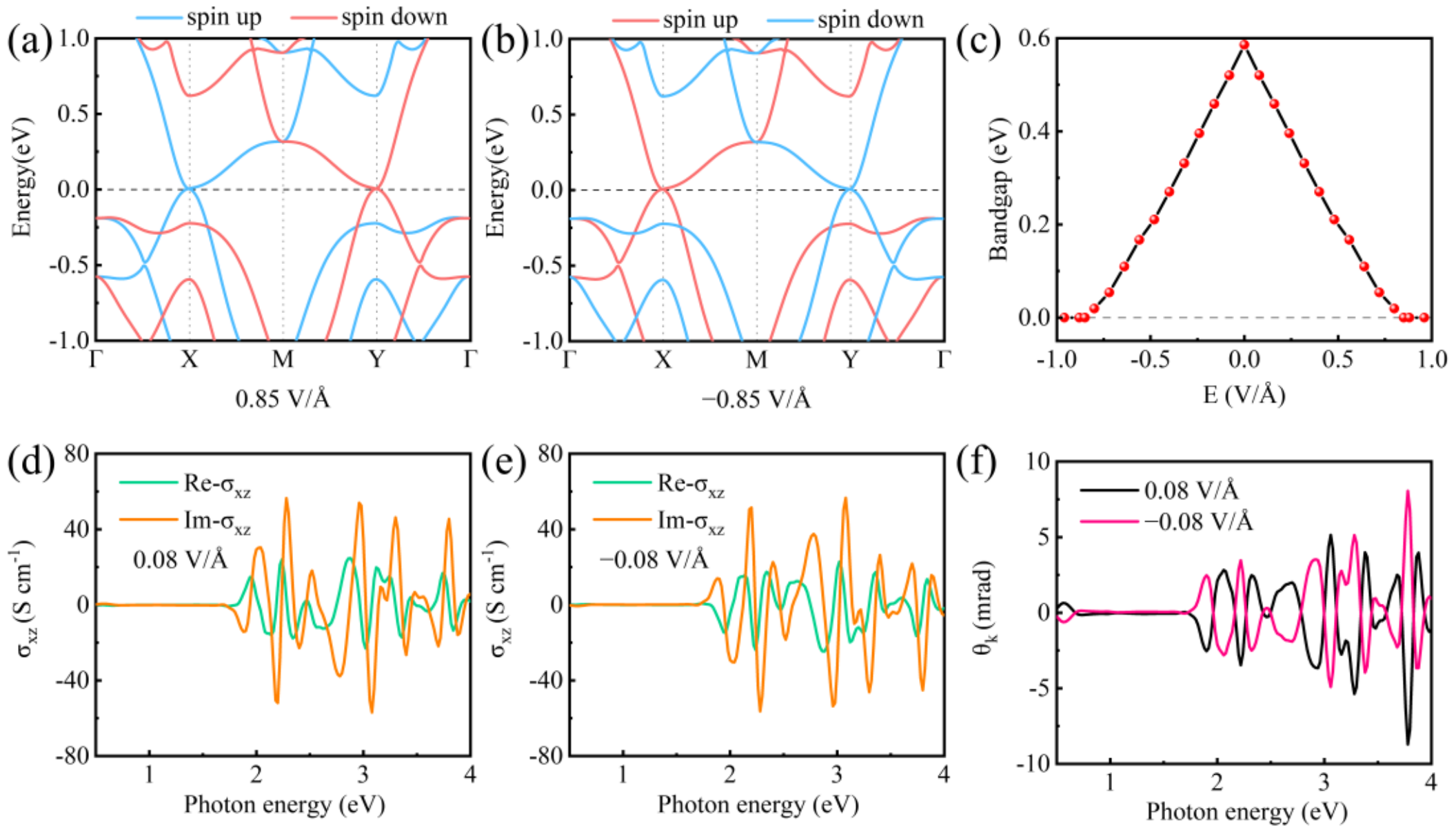

FIG. 2. Band structures of bilayer $Cr_2SeO$ under out-of-plane electric fields of (a) 0.85 V/Å and (b) −0.85 V/Å. (c) Electric-field dependence of the band gap in bilayer $Cr_2SeO$. Off-diagonal optical conductivity $\sigma_{xz}$ under out-of-plane electric fields of (d) 0.08 V/Å and (e) −0.08 V/Å, with the green and orange lines representing the real and imaginary parts of $\sigma_{xz}$, respectively. (f) Magneto-optical Kerr rotation $\theta_K$ spectra under out-of-plane electric fields of ±0.08 V/Å.

## B. Magneto-optical Kerr effect

The MOKE provides a sensitive, noncontact probe of magnetic order through the symmetry-dependent optical response in reflection [51-53]. Whether a Kerr response is symmetry-allowed is determined by the magnetic point group. With the Néel vector oriented along the $y$ direction, bilayer $Cr_2SeO$ belongs to the magnetic point group $m'mm$, which forbids the off-diagonal optical conductivity components responsible for Kerr rotation. Therefore, bilayer $Cr_2SeO$ exhibits no MOKE at zero electric field. An out-of-plane electric field breaks the symmetry relating the two layers and lowers the magnetic point group to $m'm2'$. Consequently, nonzero off-diagonal components of the optical conductivity tensor become symmetry-allowed, thereby enabling the otherwise forbidden magneto-optical response. For the $m'm2'$ magnetic point group, the optical conductivity tensor relevant to the longitudinal MOKE takes the form [53,54]:

$$\sigma(m'm2') = \begin{pmatrix} \sigma_{xx} & 0 & \sigma_{xz} \\ 0 & \sigma_{yy} & 0 \\ -\sigma_{xz} & 0 & \sigma_{zz} \end{pmatrix}. \tag{3}$$

The complex Kerr angle can be obtained from the optical conductivity as [55,56]:

$$\phi_K(L) = \theta_K + i\eta_K = -\frac{\sigma_{xz}}{\sigma_{xx}\sqrt{1 + \frac{4\pi i}{\omega}\sigma_{xx}}}, \tag{4}$$

where the real ($\theta_K$) and imaginary ($\eta_K$) parts correspond to the Kerr rotation angle and Kerr ellipticity. Figs. 2(d) and 2(e) show the off-diagonal components of the photon-energy-dependent optical conductivity of bilayer $Cr_2SeO$ under out-of-plane electric fields of 0.08 V/Å and −0.08 V/Å, respectively. Their nonzero values demonstrate the emergence of a magneto-optical response under the applied electric field. The longitudinal Kerr rotation spectra of bilayer $Cr_2SeO$ under opposite out-of-plane electric fields are shown in Fig. 2(f). The maximum Kerr rotation angle is approximately 8.06 mrad in the ultraviolet region, while several peaks also appear in the visible range. At a given field magnitude, reversing the electric-field direction changes the sign of both the real and imaginary parts of $\sigma_{xz}$, thereby reversing the sign of the Kerr rotation angle. Reversing the electric-field direction simultaneously reverses the spin-momentum-locking pattern and the sign of the Kerr rotation angle. Therefore, the MOKE provides an optical signature of electric-field-controlled spin-momentum locking.

### C. Anisotropic spin transport

The electric-field-induced semiconductor-to-metal transition provides conducting states at the Fermi level in bilayer $Cr_2SeO$. In addition, the electronic states near the high-symmetry points X and Y exhibit pronounced spin polarization at the Fermi level, and the spin-polarization directions of these states reverse when the electric-field direction is switched. These features suggest that the spin-dependent transport can be controlled electrically and may exhibit pronounced directional dependence. We therefore calculate the spin-resolved charge conductivity within Boltzmann transport theory. Here, an out-of-plane electric field is used to tune the electronic states of the bilayer, while an in-plane electric field serves as the driving force for transport. Fig. 3(a) presents the energy-dependent spin-resolved conductivity $\sigma_{xx}$ along the $x$ direction for bilayer $Cr_2SeO$ under an out-of-plane electric field of 0.85 V/Å. The two spin channels exhibit markedly different energy dependences, with their conductivity spectra clearly separated near the Fermi level.

To further investigate the anisotropic spin transport, we calculate the longitudinal $\sigma_L^{\uparrow,\downarrow}$ and transverse $\sigma_T^{\uparrow,\downarrow}$ spin-resolved charge conductivities as functions of the angle $\theta$, where $\theta$ denotes the angle between the in-plane driving electric field and the $x$ axis. As shown in Fig. 3(b), the spin-resolved longitudinal and transverse charge conductivities exhibit a pronounced twofold angular dependence with a period of π. The longitudinal conductivities of the two spin channels vary oppositely with $\theta$, while the transverse components have equal magnitudes but opposite signs. The total transverse charge conductivity remains zero for all $\theta$ because the transverse conductivities of the two spin channels cancel exactly. By contrast, the transverse spin conductivity is finite for general in-plane electric-field directions and vanishes only when the electric field is aligned with the $x$ or $y$ axis. The corresponding longitudinal and transverse spin conductivities, defined as $\sigma_L^S = \sigma_L^{\uparrow} - \sigma_L^{\downarrow}$ and $\sigma_T^S = \sigma_T^{\uparrow} - \sigma_T^{\downarrow}$, respectively, are shown in Fig. 3(c). Their pronounced angular dependence shows that the longitudinal and transverse spin conductivities can be continuously tuned by rotating the in-plane electric field. When the out-of-plane electric field is reversed from 0.85 to −0.85 V/Å, the spin-up and spin-down conductivity curves are exchanged, as shown in Fig. 3(d). The angular dependences of the spin-resolved longitudinal and transverse charge conductivities remain π-periodic, while the spin-up and spin-down components are interchanged, as shown in Fig. 3(e). Consequently, the longitudinal and transverse spin conductivities reverse sign for a fixed orientation of the in-plane driving field, as shown in Fig. 3(f). The origin of the electrically controlled spin transport lies in the electric-field-controlled spin-momentum-locking pattern. Switching the direction of the electric field reverses the spin polarization of the bands, thereby exchanging the spin-up and spin-down conducting channels and reversing the signs of the longitudinal and transverse spin conductivities. These results demonstrate effective electric-field control of spin transport in bilayer $Cr_2SeO$ without reversing the Néel vector.

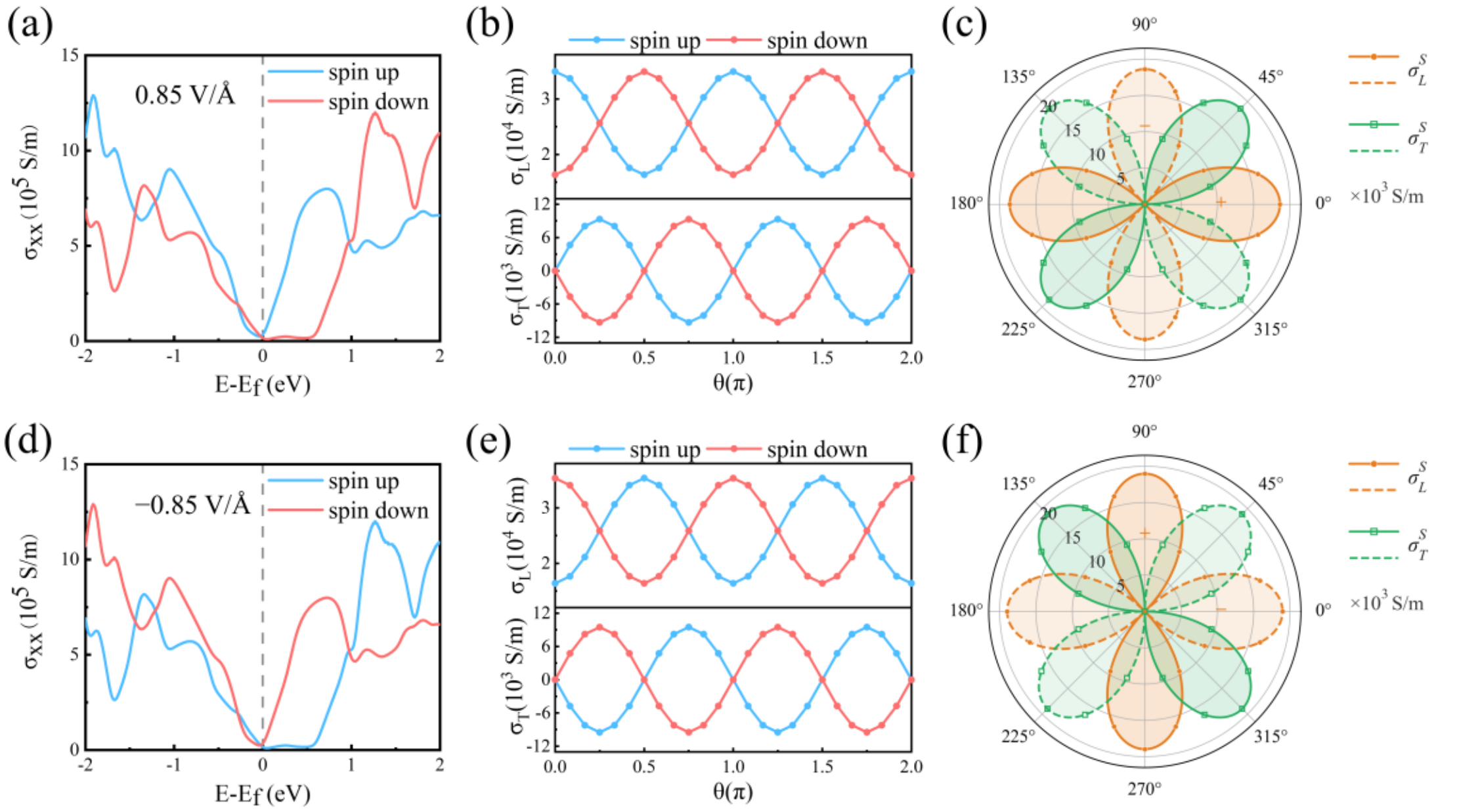

FIG. 3. Spin transport properties of bilayer $Cr_2SeO$ under out-of-plane electric fields of (a)–(c) 0.85 V/Å and (d)–(f) −0.85 V/Å. The energy-dependent spin-resolved charge conductivities along the $x$ direction $\sigma_{xx}$ are shown in (a) and (d). The angle-dependent longitudinal $\sigma_L^{\uparrow,\downarrow}$ and transverse $\sigma_T^{\uparrow,\downarrow}$ spin-resolved charge conductivities are shown in (b) and (e). The corresponding longitudinal $\sigma_L^S$ and transverse $\sigma_T^S$ spin conductivities are shown in (c) and (f). Blue and red denote the spin-up and spin-down channels, respectively, while orange and green denote the longitudinal and transverse spin conductivities, respectively. Solid and dashed portions of the curves in (c) and (f) denote positive and negative values of the spin conductivities, respectively.

## D. All-electrically controlled altermagnetic tunnel junction

Taking advantage of the electric-field-controlled spin-momentum locking in bilayer $Cr_2SeO$, we propose a dual-gate bilayer $Cr_2SeO$ AMTJ without reversing the Néel vector, as shown in Fig. 4(a). In contrast to previously reported van der Waals-stacked AMTJs, our device employs an in-plane transport geometry and consists entirely of bilayer $Cr_2SeO$. Two independently controlled top gates are used to tune the left and right electrode regions, including their adjacent buffer layers. The left electrode is subjected to a fixed gate voltage $+V_g$, while the gate voltage applied to the right electrode can be chosen as $+V_g$ or $-V_g$. The tunneling barrier consists of N tetragonal unit cells of bilayer $Cr_2SeO$. When the gate voltages applied to the left and right electrodes have the same sign, the AMTJ is in the PC state. When they have opposite signs, the AMTJ is in the APC state. The PC state exhibits low resistance, whereas the APC state exhibits high resistance. The resistance states discussed here are induced and maintained by the applied gate fields.

As shown in Figs. 4(b) and 4(c), we use the qualitative 2D Fermi surfaces to illustrate the operational principle of the bilayer $Cr_2SeO$ AMTJ. In the PC state, the spin-polarized conducting channels at the Fermi level are spin-matched between the left and right electrodes at the same **k** in momentum space. Due to quantum tunneling effects, both spin-up and spin-down electrons can transmit through the AMTJ, resulting in a low-resistance state. In contrast, in the APC state, these channels are spin-mismatched between the two electrodes at the same **k**. Owing to the tunneling barrier and the spin mismatch between the conducting channels, the transmission of both spin channels is strongly suppressed, leading to a high-resistance state. By independently controlling the spin-momentum locking in the two electrodes through the applied gate voltages, the spin-polarized conducting channels can therefore be switched between spin-matched and spin-mismatched configurations at the same **k**, enabling efficient electrical switching between the low- and high-resistance states.

To assess the feasibility of the bilayer $Cr_2SeO$ AMTJ, we employ the DFT-NEGF method to calculate its electronic transport properties with a tunneling barrier consisting of $N = 3$ bilayer $Cr_2SeO$ unit cells. Figs. 4(d) and 4(e) presents the zero-bias transmission spectra for the PC and APC states at $V_g = 13\ V$, respectively. Here, $V_g = 13\ V$ is chosen based on the critical voltage for the semiconductor-to-metal transition obtained from calculations performed using the Nanodcal package. In the PC state, both the spin-up and spin-down channels exhibit finite transmission near the Fermi level, indicating that both spin channels can contribute to the current under a small applied bias. In contrast, the transmission of both spin channels in the APC state is strongly suppressed near the Fermi level, indicating that neither spin channel can contribute appreciably to the current under a small applied bias and thus giving rise to a high-resistance state. The pronounced contrast between the PC and APC transmission spectra validates the proposed dual-gate control mechanism, demonstrating that a tunneling magnetoresistance effect can be realized in the AMTJ solely through electric-field control, without reversing the Néel vector.

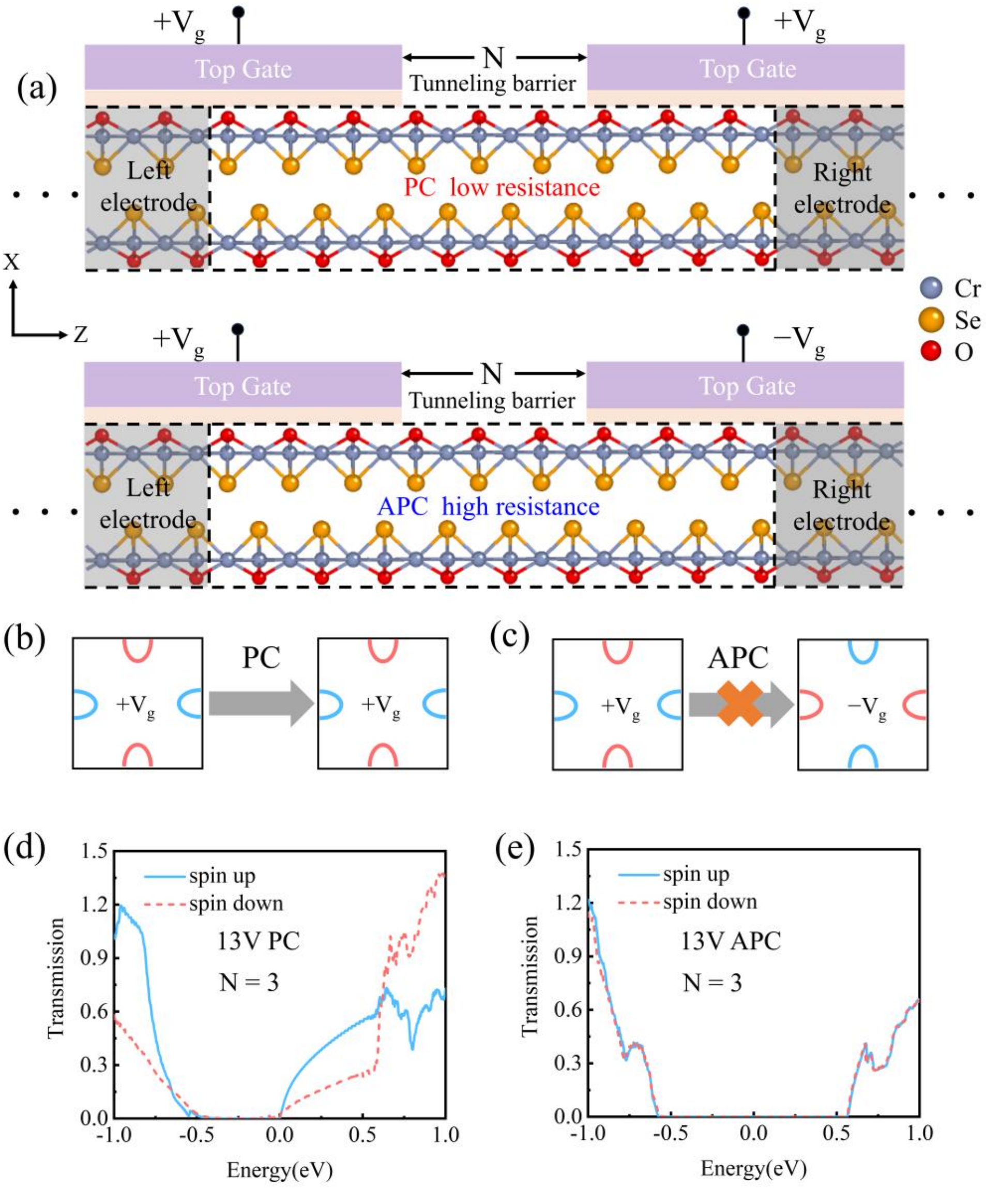

FIG. 4. (a) Schematic diagram of the bilayer $Cr_2SeO$ AMTJ for $N = 3$. Independent top gates apply out-of-plane electric fields to the left and right electrode regions and their adjacent buffer layers. The central region, which is not covered by either top gate, serves as the tunneling barrier. The qualitative 2D Fermi surfaces of the left and right electrodes in the (b) PC and (c) APC states. Red and blue denote opposite spin channels. Spin-resolved transmission spectra of the bilayer $Cr_2SeO$ AMTJ for $N = 3$ and $V_g = 13\,V$ at zero bias in the (d) PC and (e) APC states, respectively.

To further study the device performance under nonequilibrium conditions, we calculate its TMR under finite bias voltage for $V_g = 13\,V$, as illustrated in Fig. 5. The TMR is defined as $\mathrm{TMR} = (I_{PC} - I_{APC})/I_{APC}$, where $I_{PC}$ and $I_{APC}$ are the total currents in the PC and APC states, respectively. All the devices considered here exhibit robust and ultrahigh TMR over the finite-bias range. For $N = 3$, the TMR remains above $10^5$ within a bias window of −0.08 to 0.08 V, whereas for $N = 4$ and $N = 5$, it remains above $10^5$ throughout the investigated bias range of −0.10 to 0.10 V. Moreover, the maximum TMR for $N = 4$ and $N = 5$ reaches the order of $10^8$, demonstrating the excellent magnetoresistive performance of the proposed AMTJ. Ultrahigh TMR is also maintained for different tunneling barrier lengths. These results demonstrate that the bilayer $Cr_2SeO$ AMTJ enables electrically controlled ultrahigh TMR, with values substantially exceeding those of most previously reported 2D and conventional three-dimensional MTJs [24,31,34,57-59], highlighting its potential for spintronic applications.

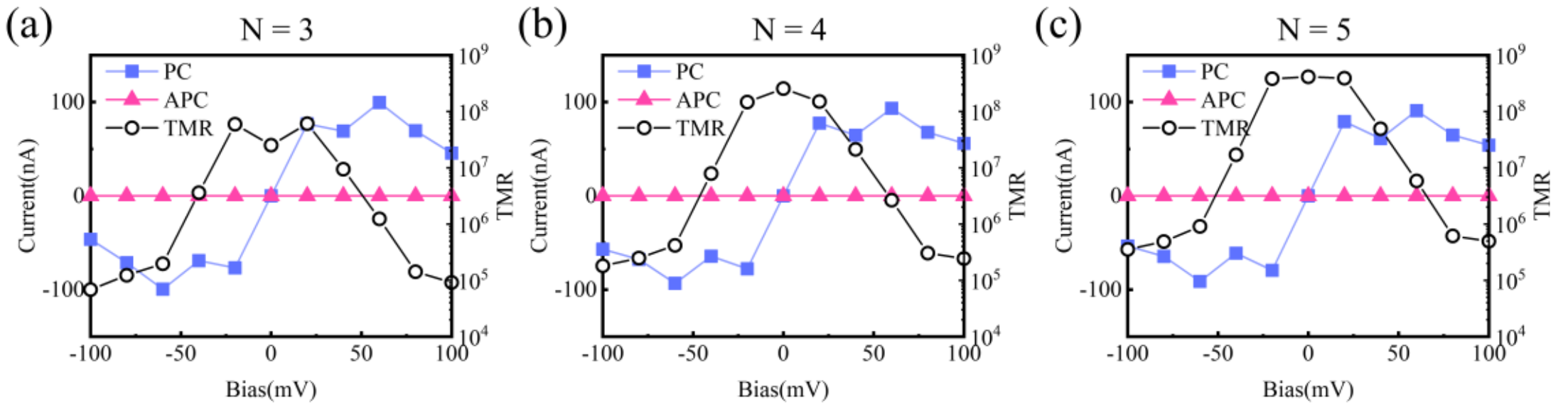

FIG. 5. Bias-dependent currents in the PC and APC states (left axis) and the corresponding TMR (right axis) of the bilayer $Cr_2SeO$ AMTJ at $V_g = 13$ V for (a) N = 3, (b) N = 4, and (c) N = 5. At zero bias, the TMR is evaluated from the transmission coefficients rather than the current.

In summary, we have demonstrated that an out-of-plane electric field lifts the spin degeneracy of bilayer $Cr_2SeO$ and induces spin-momentum locking. Reversing the electric-field direction reverses the spin-momentum-locking pattern and the sign of the Kerr rotation angle, while a sufficiently strong field drives a semiconductor-to-metal transition. The electric field also enables the modulation of the anisotropic transport properties of bilayer $Cr_2SeO$. Building on the electric-field-controlled spin-momentum locking, we propose a dual-gate bilayer $Cr_2SeO$ AMTJ in which the high- and low-resistance states are switched by changing the relative orientations of the gate voltages applied to the two electrodes, without reversing the Néel vector. The maximum TMR ratio reaches the order of $10^8$, while the TMR ratio remains above $10^5$ over a broad bias-voltage window. The proposed dual-gate scheme can be extended to other bilayers that exhibit similarly electrically controllable spin-momentum locking. Our work provides a theoretical basis for developing all-electrically controlled altermagnetic devices.

## ACKNOWLEDGMENTS

This work was funded by the Key Project of the Natural Science Program of Xinjiang Uygur Autonomous Region (Grant No. 2023D01D03), the XJU Innovation Project for Ph.D candidate (XJU2024BS049), the Tianchi-Talent Project for Young Doctors of Xinjiang Uygur Autonomous Region, the National Natural Science Foundation of China (Grant Nos. 52073308; 12564028), the Tianshan Innovation Team Program of Xinjiang Uygur Autonomous Region (Grant No. 2023D14001), the Natural Science Foundation of Xinjiang Uygur Autonomous Region (Grant No. 2024D01C233) and the State Key Laboratory of Powder Metallurgy at Central South University. This work was carried out in part using computing resources at the High-Performance Computing Center of Central South University.

## DATA AVAILABILITY

The data that support the findings in this manuscript are available from the authors upon reasonable request.

## REFERENCES

[1] H.-Y. Ma, M. Hu, N. Li, J. Liu, W. Yao, J.-F. Jia, and J. Liu, Multifunctional antiferromagnetic materials with giant piezomagnetism and noncollinear spin current, Nat. Commun. **12**, 2846 (2021).

[2] B. Pan, P. Zhou, P. Lyu, H. Xiao, X. Yang, and L. Sun, General Stacking Theory for Altermagnetism in Bilayer Systems, Phys. Rev. Lett. **133**, 166701 (2024).

[3] W. Sun, W. Wang, C. Yang, R. Hu, S. Yan, S. Huang, and Z. Cheng, Altermagnetism Induced by Sliding Ferroelectricity via Lattice Symmetry-Mediated Magnetoelectric Coupling, Nano Lett. **24**, 11179 (2024).

[4] S.-D. Guo, X.-S. Guo, K. Cheng, K. Wang, and Y. S. Ang, Piezoelectric altermagnetism and spin-valley polarization in Janus monolayer $Cr_2SO$, Appl. Phys. Lett. **123**, 082401 (2023).

[5] Z. Liu, M. Ozeki, S. Asai, S. Itoh, and T. Masuda, Chiral Split Magnon in Altermagnetic MnTe, Phys. Rev. Lett. **133**, 156702 (2024).

[6] G. Yang, Z. Li, S. Yang, J. Li, H. Zheng, W. Zhu, Z. Pan, Y. Xu, S. Cao, W. Zhao *et al.*, Three-dimensional mapping of the altermagnetic spin splitting in CrSb, Nat. Commun. **16**, 1442 (2025).

[7] Y. Wu, L. Deng, X. Yin, J. Tong, F. Tian, and X. Zhang, Valley-Related Multipiezo Effect and Noncollinear Spin Current in an Altermagnet $Fe_2Se_2O$ Monolayer, Nano Lett. **24**, 10534 (2024).

[8] R.-W. Zhang, C. Cui, R. Li, J. Duan, L. Li, Z.-M. Yu, and Y. Yao, Predictable Gate-Field Control of Spin in Altermagnets with Spin-Layer Coupling, Phys. Rev. Lett. **133**, 056401 (2024).

[9] R. Bhattarai, P. Minch, and T. D. Rhone, High-throughput screening of altermagnetic materials, Phys. Rev. Mater. **9**, 064403 (2025).

[10] F. Zhang, X. Cheng, Z. Yin, C. Liu, L. Deng, Y. Qiao, Z. Shi, S. Zhang, J. Lin, Z. Liu *et al.*, Crystal-symmetry-paired spin–valley locking in a layered room-temperature metallic altermagnet candidate, Nat. Phys. **21**, 760 (2025).

[11] R. Xu, Y. Gao, and J. Liu, Chemical design of monolayer altermagnets, Natl. Sci. Rev. **13**, nwaf528 (2026).

[12] I. Khan, D. Bezzerga, B. Marfoua, and J. Hong, Altermagnetism, piezovalley, and ferroelectricity in two-dimensional $Cr_2SeO$ altermagnet, npj 2D Mater. Appl. **9**, 18 (2025).

[13] K. Samanta, D.-F. Shao, and E. Y. Tsymbal, Spin Filtering with Insulating Altermagnets, Nano Lett. **25**, 3150 (2025).

[14] X. Zhou, W. Feng, R.-W. Zhang, L. Šmejkal, J. Sinova, Y. Mokrousov, and Y. Yao, Crystal

Thermal Transport in Altermagnetic $RuO_2$, Phys. Rev. Lett. **132**, 056701 (2024).
[15] M. Gu, Y. Liu, H. Zhu, K. Yananose, X. Chen, Y. Hu, A. Stroppa, and Q. Liu, Ferroelectric Switchable Altermagnetism, Phys. Rev. Lett. **134**, 106802 (2025).
[16] Y.-Q. Li, Y.-K. Zhang, X.-L. Lu, Y.-P. Shao, Z.-Q. Bao, J.-D. Zheng, W.-Y. Tong, and C.-G. Duan, Ferrovalley Physics in Stacked Bilayer Altermagnetic Systems, Nano Lett. **25**, 6032 (2025).
[17] X. Duan, J. Zhang, Z. Zhu, Y. Liu, Z. Zhang, I. Žutić, and T. Zhou, Antiferroelectric Altermagnets: Antiferroelectricity Alters Magnets, Phys. Rev. Lett. **134**, 106801 (2025).
[18] S. Iguchi, H. Kobayashi, Y. Ikemoto, T. Furukawa, H. Itoh, S. Iwai, T. Moriwaki, and T. Sasaki, Magneto-optical spectra of an organic antiferromagnet as a candidate for an altermagnet, Phys. Rev. Res. **7**, 033026 (2025).
[19] M. L. Liu, J. Sun, and S. Meng, Uncompensated Linear Dichroism of Magneto-Optical Kerr Effect in a 2D Altermagnet, Adv. Funct. Mater. **36** (2026).
[20] S. Wu, P. Diao, W. Sun, C. Yang, S. Huang, and Z. Cheng, Band-Rearrangement-Enabled Nonvolatile Altermagnetism Switching in 2D Ferroelectric/Magnetic Heterostructures, Adv. Funct. Mater. **36**, e16174 (2026).
[21] X. Zhang, P. Jiang, L.-Y. Xu, L. Wang, L. Liu, H.-M. Huang, T. Cao, and Y.-L. Li, Giant Spin Splitting and Anisotropic Spin Polarization in 2D Altermagnet $Cr_2O$, Nano Lett. **25**, 16547 (2025).
[22] Q. Wu and L. K. Ang, Giant tunneling magnetoresistance in atomically thin $VSi_2N_4/MoSi_2N_4/VSi_2N_4$ magnetic tunnel junction, Appl. Phys. Lett. **120**, 022401 (2022).
[23] F. Li, B. Yang, Y. Zhu, X. Han, and Y. Yan, Ultrahigh tunneling magnetoresistance in van der Waals and lateral magnetic tunnel junctions formed by intrinsic ferromagnets $Li_{0.5}CrI_3$ and $CrI_3$, Appl. Phys. Lett. **117**, 022412 (2020).
[24] E. Balcı, Ü. Ö. Akkuş, and S. Berber, High TMR in MXene-Based $Mn_2CF_2/Ti_2CO_2/Mn_2CF_2$ Magnetic Tunneling Junction, ACS Appl. Mater. Interfaces **11**, 3609 (2019).
[25] X. Zhang, P. Gong, F. Liu, J. Wu, and S. Zhu, $Mn_2NT_2$ (T = O, F) Nanosheets for Electrically and Thermally Driven Magnetic Tunnel Junctions, ACS Appl. Nano Mater. **5**, 15183 (2022).
[26] H. Yu, M. Chen, Z. Shao, Y. Tao, X. Jiang, Y. Dong, J. Zhang, X. Yang, and Y. Liu, Giant tunneling magnetoresistance in in-plane double-barrier magnetic tunnel junctions based on MXene $Cr_2C$, Phys. Chem. Chem. Phys. **25**, 10991 (2023).
[27] Q. Chen, X. Zheng, P. Jiang, Y.-H. Zhou, L. Zhang, and Z. Zeng, Electric field induced tunable half-metallicity in an A-type antiferromagnetic bilayer $LaBr_2$, Phys. Rev. B **106**, 245423 (2022).
[28] Q. Lu, W.-J. Gong, S. Li, X.-T. Zu, and H.-F. Lü, Half-metallic CrAs nanosheet for magnetic tunnel junctions, Phys. Rev. Appl. **22**, 024002 (2024).
[29] Q.-Q. Li, Z.-F. Duan, W.-W. Liu, R. Yang, B. Li, and K.-Q. Chen, Two-dimensional lateral magnetic tunnel junction with ultrahigh tunneling magnetoresistance, Nano Res. **18**, 94907188 (2025).
[30] Y. Feng, X. Wu, J. Han, and G. Gao, Robust half-metallicities and perfect spin transport properties in 2D transition metal dichlorides, J. Mater. Chem. C **6**, 4087 (2018).
[31] D. Waldron, V. Timoshevskii, Y. Hu, K. Xia, and H. Guo, First Principles Modeling of Tunnel Magnetoresistance of Fe/MgO/Fe Trilayers, Phys. Rev. Lett. **97**, 226802 (2006).
[32] S. Yuasa, T. Nagahama, A. Fukushima, Y. Suzuki, and K. Ando, Giant room-temperature magnetoresistance in single-crystal Fe/MgO/Fe magnetic tunnel junctions, Nat. Mater. **3**, 868 (2004).
[33] B. Dieny, I. L. Prejbeanu, K. Garello, P. Gambardella, P. Freitas, R. Lehndorff, W. Raberg, U. Ebels, S. O. Demokritov, J. Akerman *et al.*, Opportunities and challenges for spintronics in the microelectronics industry, Nat. Electron. **3**, 446 (2020).

[34] J. Wang, X. Yang, Z. Yang, J. Lu, P. Ho, W. Wang, Y. S. Ang, Z. Cheng, and S. Fang, Pentagonal 2D Altermagnets: Material Screening and Altermagnetic Tunneling Junction Device Application, Adv. Funct. Mater. **36**, 2505145 (2026).

[35] J.-J. He, L.-X. Liu, Y.-D. Guo, J.-R. Yuan, X.-H. Yan, and S. Sanvito, Tunnel-like transport and extremely large magnetoresistance in an all-metal junction incorporating the altermagnet $KV_2Se_2O$, Phys. Rev. B **113**, L060410 (2026).

[36] S. Sheoran, L. Keenan, D. Nell, and S. Sanvito, Negative Differential Resistance and Ultrahigh TMR in Altermagnetic Tunnel Junctions, Nano Lett. **26**, 10329 (2026).

[37] L. Zhang, Y. Yan, X. Wu, S. Azam, and G. Gao, Altermagnetic-$RuO_2$-based all-magnetic tunnel junction: Giant and multistate tunneling magnetoresistances with perfect spin filtering, Appl. Phys. Lett. **128**, 202403 (2026).

[38] C. Mao, S. Liu, B. Wu, S. Li, J. Yang, and J. Yang, Electrical-controlled and layer-filtered altermagnetic tunnel junction with all-in-one architecture, Appl. Phys. Lett. **129**, 062401 (2026).

[39] L. Han, X. Fu, R. Peng, X. Cheng, J. Dai, L. Liu, Y. Li, Y. Zhang, W. Zhu, H. Bai *et al.*, Electrical 180° switching of Néel vector in spin-splitting antiferromagnet, Sci. Adv. **10**, eadn0479.

[40] Z. Zhou, X. Cheng, M. Hu, R. Chu, H. Bai, L. Han, J. Liu, F. Pan, and C. Song, Manipulation of the altermagnetic order in CrSb via crystal symmetry, Nature **638**, 645 (2025).

[41] J. Tian, J. Li, H. Liu, Y. Li, Z. Liu, L. Li, J. Li, G. Liu, and J. Shi, Spin-layer coupling in an altermagnetic multilayer: A design principle for spintronics, Phys. Rev. B **111**, 035437 (2025).

[42] X. Chen, X. Xuan, W. Guo, and Z. Zhang, Cross-State Alternating Magnetism in Two-Dimensional Systems, Nano Lett. **25**, 18068 (2025).

[43] G. Kresse and J. Furthmüller, Efficiency of ab-initio total energy calculations for metals and semiconductors using a plane-wave basis set, Comput. Mater. Sci. **6**, 15 (1996).

[44] P. E. Blöchl, Projector augmented-wave method, Phys. Rev. B **50**, 17953 (1994).

[45] J. P. Perdew, K. Burke, and M. Ernzerhof, Generalized Gradient Approximation Made Simple, Phys. Rev. Lett. **77**, 3865 (1996).

[46] S. Grimme, Semiempirical GGA-type density functional constructed with a long-range dispersion correction, J. Comput. Chem. **27**, 1787 (2006).

[47] A. A. Mostofi, J. R. Yates, Y.-S. Lee, I. Souza, D. Vanderbilt, and N. Marzari, wannier90: A tool for obtaining maximally-localised Wannier functions, Comput. Phys. Commun. **178**, 685 (2008).

[48] J. R. Yates, X. Wang, D. Vanderbilt, and I. Souza, Spectral and Fermi surface properties from Wannier interpolation, Phys. Rev. B **75**, 195121 (2007).

[49] G. Pizzi, D. Volja, B. Kozinsky, M. Fornari, and N. Marzari, BoltzWann: A code for the evaluation of thermoelectric and electronic transport properties with a maximally-localized Wannier functions basis, Comput. Phys. Commun. **185**, 422 (2014).

[50] J. Taylor, H. Guo, and J. Wang, Ab initio modeling of quantum transport properties of molecular electronic devices, Phys. Rev. B **63**, 245407 (2001).

[51] K. F. Mak, J. Shan, and D. C. Ralph, Probing and controlling magnetic states in 2D layered magnetic materials, Nat. Rev. Phys. **1**, 646 (2019).

[52] N. Wang, J. Chen, N. Ding, H. Zhang, S. Dong, and S.-S. Wang, Magneto-optical Kerr effect and magnetoelasticity in a weakly ferromagnetic $RuF_4$ monolayer, Phys. Rev. B **106**, 064435 (2022).

[53] Y. Yu, Z. Zhang, H. Ye, D. Zhang, L. Chen, G. P. Zhang, and J. Wang, Strain-tunable magnetic anisotropy and magneto-optical effects in a room-temperature $Cr_2Te_3$ monolayer spin valve, Phys. Rev. B **112**, 144431 (2025).

[54] K. Yang, W. Hu, H. Wu, M.-H. Whangbo, P. G. Radaelli, and A. Stroppa, Magneto-Optical Kerr Switching Properties of $(CrI_3)_2$ and $(CrBr_3/CrI_3)$ Bilayers, ACS Appl. Electron. Mater. **2**, 1373 (2020).
[55] D. Sangalli, A. Marini, and A. Debernardi, Pseudopotential-based first-principles approach to the magneto-optical Kerr effect: From metals to the inclusion of local fields and excitonic effects, Phys. Rev. B **86**, 125139 (2012).
[56] B. Gu, S. Takahashi, and S. Maekawa, Enhanced magneto-optical Kerr effect at Fe/insulator interfaces, Phys. Rev. B **96**, 214423 (2017).
[57] W. Jin, G. Zhang, H. Wu, L. Yang, W. Zhang, and H. Chang, Room-Temperature and Tunable Tunneling Magnetoresistance in $Fe_3GaTe_2$-Based 2D van der Waals Heterojunctions, ACS Appl. Mater. Interfaces **15**, 36519 (2023).
[58] Z.-W. Zhang, Y.-F. Lang, H.-P. Zhu, B. Li, Y.-Q. Zhao, B. Wei, and W.-X. Zhou, Electronic structure, Curie temperature, and magnetic transport of a two-dimensional multiferroic $MnSe_2/In_2Se_3$ heterostructure, Phys. Rev. Appl. **21**, 064012 (2024).
[59] H.-x. Liu, Y. Honda, T. Taira, K.-i. Matsuda, M. Arita, T. Uemura, and M. Yamamoto, Giant tunneling magnetoresistance in epitaxial $Co_2MnSi/MgO/Co_2MnSi$ magnetic tunnel junctions by half-metallicity of Co2MnSi and coherent tunneling, Appl. Phys. Lett. **101**, 132418 (2012).